\documentstyle[prl,aps,epsf]{revtex}

\begin{document}

\advance\textheight by 0.2in

\draft
\twocolumn[\hsize\textwidth\columnwidth\hsize\csname@twocolumnfalse
\endcsname  

\title{Disorder Driven Melting of the Vortex Line Lattice
}

\author{P. Olsson$^1$ and S. Teitel$^2$}

\address{$^1$Department of Theoretical Physics, Ume{\aa} University, 901 87 
Ume{\aa} Sweden\\$^2$Department of Physics and Astronomy, University 
of Rochester, Rochester, NY 14627}

\date{\today}

\maketitle

\begin{abstract}

We use Monte Carlo simulations of the 3D uniformly frustrated XY
model, with uncorrelated quenched randomness in the in-plane couplings, to
model the effect of random point pins on the vortex line phases of a
type II superconductor.  We map out the phase diagram as a function of
temperature $T$ and randomness strength $p$ for fixed applied
magnetic field.
We find that, as $p$ increases to a critical value $p_c$, the first order
vortex lattice melting line turns parallel to the $T$ axis, and
continues smoothly down to low temperature, rather than ending at a
critical point.  The entropy jump across this line at $p_c$ vanishes,
but the transition remains first order.  Above this disorder driven
transition line, we find that the helicity modulus parallel to the applied
field vanishes, and so no true phase coherent vortex glass exists.

\end{abstract}


]
Experimental \cite{R1,R2,R3}, theoretical \cite{R4,R5,R6} and numerical 
\cite{R7,R8} studies have argued that
the effect of intrinsic point impurities on otherwise 
clean single crystal samples of high $T_c$ superconductors leads
to a $H-T$ phase diagram with the following generic form.
At low magnetic 
fields $H$, an elastically distorted vortex lattice (the ``Bragg 
glass'' \cite{R4}) undergoes a first order melting transition to a vortex 
liquid as temperature $T$ is increased.  This melting 
line $T_c(H)$ continues as $H$ is increased, until an ``upper critical point,''
$T_{\rm ucp}$, is reached above which sharp discontinuities in measured quantities 
become smeared.  Upon increasing 
$H$ at lower temperatures, $T<T_{\rm ucp}$,  
the vortex lattice undergoes a transformation to a disordered 
vortex state along a line, $H_{\rm sp}(T)$, characterized by the ``second 
magnetization peak'' where critical currents show a sharp increase.  
As $T$ increases, the $H_{\rm sp}(T)$ line continues to the vicinity of 
$T_{\rm ucp}$.  In BSCCO, $H_{\rm sp}(T)$ is only weakly dependent on $T$ 
\cite{R2}.
Very recently, several experiments \cite{R9} have provided 
strong evidence that the $H_{\rm sp}(T)$ line in BSCCO is associated with a  
thermodynamic first order phase transition.
Whether the disordered state above $H_{\rm sp}(T)$ is a 
``vortex glass'' \cite{R5}, characterized by true superconducting phase 
coherence and separated from the vortex liquid by a sharp phase 
transition, or whether it is a dynamically frozen state that smoothly
crosses over to the vortex liquid, remains a topic of controversy 
\cite{R10}.

Since many of the experimental and numerical studies focus on 
dynamical probes, from which it can sometimes be difficult to infer a 
true equilibrium phase transition, and analytical models must
resort to Lindemann or other simplifying approximations, it is 
important to establish
the true equilibrium phase diagram within a realistic model system.
Towards this end we have carried out extensive Monte 
Carlo (MC) studies of the uniformly frustrated three dimensional (3D) XY 
model \cite{R11}, with uncorrelated quenched random couplings to model point 
impurities.
Applying a fixed magnetic field $B$ we map out the phase diagram as a 
function of disorder strength $p$ and temperature $T$.  Increasing $p$
at fixed $B$ is believed to play a similar role as the more 
physical case of increasing $B$ at fixed $p$.  Taking great care to
achieve proper equilibration, and measuring thermodynamic derivatives 
of the free energy, we find in our model a 
single first order phase boundary.  At small $p$, the 
transition, $T_c(p)$, is a thermally driven melting of the vortex lattice.  
Increasing $p$, there is a maximum critical $p_c$ above which
disorder destroys the vortex lattice; when $p=p_c$, the 
transition line $T_c(p)$ turns parallel to the $T$ axis and smoothly 
continues down to low $T$. No 
critical point is found to separate the thermal from the disorder 
driven sections of the phase boundary.  For $p>p_c$ we find no 
evidence for a true phase coherent vortex glass.

The model we study is given by the Hamiltonian
\begin{equation}
{\cal H}[\theta_i]=-\sum_{{\rm bonds}\,i\mu}J_{i\mu}\cos
(\theta_i-\theta_{i+\hat\mu}-A_{i\mu})
\label{eH}
\end{equation}
where $\theta_i$ is the phase of the superconducting wavefunction on
site $i$ of a 3D periodic cubic grid of sites, the sum
is over all bonds in directions $\hat\mu=\hat x, \hat y, \hat z$, and
$A_{i\mu}=(2\pi/\phi_0)\int^{i+\hat\mu}_i{\bf A}\cdot{\bf d\ell}$ is the
integral of the magnetic vector potential across bond $i\mu$, where
${\bf\nabla}\times{\bf A}=B\hat z$ is a fixed uniform magnetic field
in the $\hat z$ direction.
To model uncorrelated random point vortex pinning in the $xy$ planes, we
take,
\begin{equation}
\begin{array}{ll}
    J_{i\mu}=J_z, & \mu=z  \\
    J_{i\mu}=J_\perp(1+p\epsilon_{i\mu}),\qquad & \mu=x,y
\end{array}
\label{eJ}
\end{equation}
The coupling between planes $J_z$ is uniform, while each bond in
the $xy$ plane is randomly perturbed about the constant value $J_\perp$;
the $\epsilon_{i\mu}$ are independent Gaussian random variables with
$\langle\epsilon_{\i\mu}\rangle=0$, $\langle\epsilon_{\i\mu}^2\rangle=1$.
The parameter $p$ controls the strength of the disorder.  For
computational convenience we choose $J_z/J_\perp=1/40$,
with a vortex line density of $f=a_\perp^2B/\phi_0=1/20$, where $a_\perp$
is the
grid spacing in the $xy$ plane, and $\phi_0=hc/2e$ is the flux quantum.
Our system size is $L_x=L_y=40$, with $L_z=16$.  To
check finite size effects, we have also considered $L_z=24$ and $32$
for certain cases.  Our runs are typically $1 - 10 \times 10^7$ MC
sweeps through the entire lattice near transitions.  Our results
below are for a single realization of the disorder only.  The 
extremely time consuming nature of our simulations excluded any serious 
attempt at disorder averaging.  We have, however,
carried out similar analyses for two other realizations of the 
disorder and have found qualitatively the same behavior.
\begin{figure}
\epsfxsize=3.2truein
\epsfbox{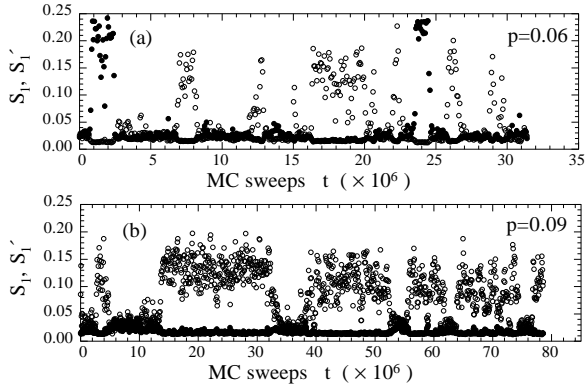}
\caption{
$S_1$ ($\bullet$) and $S_1^\prime$ ($\circ$) vs. number of Monte
Carlo sweeps $t$ 
for disorder strengths (a) $p=0.06$, $T_c=0.255$, and (b) $p=0.09$, 
$T_c=0.200$.  The 
system width is $L_z=16$.  Each data point is an average over $2^{16}$ 
sweeps.
}
\label{f1}
\end{figure}

In the pure model, $p=0$, the low temperature vortex line lattice has
a first order melting transition to a vortex line liquid \cite{R12}.  
To map out this melting transition line in the $p-T$ plane 
we fix the disorder strength $p$, and cool down from high $T$, until
we reach a temperature $T_c(p)$ at which we observe repeated hopping
back and forth between coexisting vortex liquid and lattice phases.
To detect the vortex lattice, we measure the in-plane
vortex structure function (normalized for convenience so that 
$S(0)=1$),
\begin{equation}
S({\bf k}_\perp)={1\over f^2N}\sum_{{\bf r}_\perp, z}
\langle n_z({\bf r}_\perp,z)n_z(0,z)\rangle e^{i{\bf 
k}_\perp\cdot{\bf r}_\perp}
\label{eS}
\end{equation}
where $n_z$ is the vorticity
in the $xy$ plane, and $N=L_xL_yL_z$.  
$S({\bf k}_\perp)$ will have Bragg peaks at the reciprocal lattice
vectors $\{ {\bf K}\}$ of the vortex lattice.  We
find that the vortex lattice always orders into the same periodicity
as that of the pure $p=0$ case, where there are two possible lattice
orientations related by a $90^\circ$ rotation.  
Defining $S_1$ as the
average of $S({\bf K})$ over the six smallest non-zero $\{ {\bf K}\}$ for
one lattice orientation, and $S_1^\prime$ as that for the other
orientation, we identify the vortex lattice as states in which
either $S_1$ {\it or} $S_1^\prime$ is large, according to which of
the two lattice orientations has formed; in contrast, in the
vortex liquid, both $S_1$ and $S_1^\prime$ are small.
In Fig.\,1 we plot $S_1$ and $S_1^\prime$ vs. MC simulation time $t$
at $T_c(p)$ for two disorder strengths.  For the weaker disorder,
$p=0.06$, we see both orientations of lattice coexisting with the
liquid.  For the stronger $p=0.09$, the 
disorder sufficiently breaks the degeneracy of the 
two lattice orientations, 
so that we see only coexistence between one
particular lattice orientation and the liquid.  
The repeated hopping between lattice and liquid in Fig.\,1
verifies that we are well equilibrated.  In this manner, by varying $p$, we
determine the melting phase boundary $T_c(p)$ shown in Fig.\,2.
We include in Fig.\,2 several data points for systems with $L_z=24$, 
showing only a small shift in the phase boundary as $L_z$ increased.  
\begin{figure}
\epsfxsize=3.2truein
\epsfbox{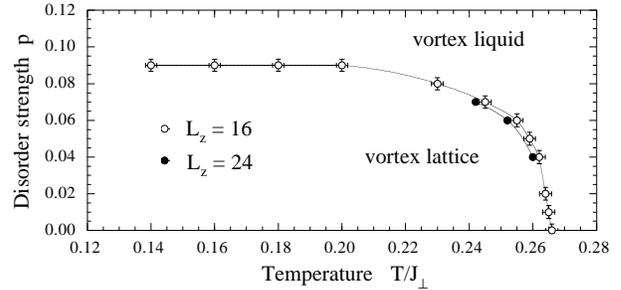}
\vspace{9pt}
\caption{
Vortex lattice melting phase boundary in the $p-T$ plane,
for $L_z=16$ ($\circ$) and $L_z=24$ ($\bullet$).
}
\label{f2}
\end{figure}

We now determine if the 
melting transition $T_c(p)$ remains 1st order, as $p$ increases.  
First order transitions are characterized by discontinuous jumps
in thermodynamic quantities.
Here we consider the 
average energy per site $E$ (the conjugate variable to temperature $T$), and a 
variable $Q$ defined to be conjugate to the disorder strength $p$,
\begin{eqnarray}
	 & E & =-{1\over N}\sum_{i\mu}J_{i\mu}\langle\cos (\theta_i
     - \theta_{i+\hat\mu}-A_{i\mu})\rangle
	\label{eq:eE}  \\
   & Q & \equiv{1\over N}{\partial F\over \partial p}=-{J_\perp\over N}
   \sum_{i,\mu=x,y}\epsilon_{i\mu}\langle\cos (\theta_i
     - \theta_{i+\hat\mu}-A_{i\mu})\rangle\enspace,
	\label{eq:eQ}
\end{eqnarray}
where $F$ is the total free energy.  

To see if there is a discrete jump in $E$ or $Q$ at $T_c(p)$, 
we use the values of $S_1$ and $S_1^\prime$ to sort microscopic
states as either vortex lattice or liquid.   We can then compute
the properties of each phase separately.
In Figs.\,3a and 3b we show semilog plots of the histograms $P(\Delta S_1)$
of values of $\Delta S_1\equiv S_1-S_1^\prime$ 
encountered during our simulation at $T_c(p)$, 
for the two cases of Fig.\,1.  In Fig.\,3a, for $p=0.06$, we see 
separated peaks for the liquid at $\Delta S_1=0$, and for the
two lattice orientations at finite 
positive and negative values of $\Delta S_1$.
In Fig.\,3b, for the stronger $p=0.09$, we see only peaks for the liquid 
and one of the two lattice orientations.  Fitting these peaks to 
empirical forms (Gaussian for the lattice, exponential for the 
liquid; these are the solid lines in Figs.\,3a,b), 
we determine the relative probability for a state with a 
given value of $\Delta S_1$ to belong to the liquid phase, or either 
of the two orientations of the lattice phases.  Sorting though 
our microscopic states we probabilistically assign each to one 
of these three phases.  We then plot 
the histograms of $E$ and $Q$ values separately for each phase.  
\begin{figure}
\epsfxsize=3.2truein
\epsfbox{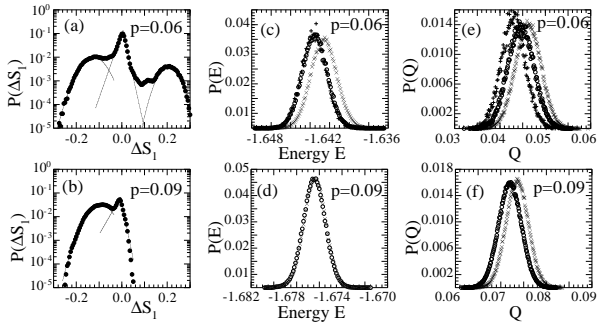}
\caption{Histograms of $\Delta S_1\equiv S_1-S_1^\prime$, energy $E$, 
and disorder conjugate $Q$ for $p=0.06$, $T_c=0.255$ and 
$p=0.09$, $T_c=0.200$, for $L_z=16$,
at the melting temperature $T_c(p)$.  ($\circ$) and ($+$) are for the 
two lattice orientations, ($\times$) are for the liquid.
}
\label{f3}
\end{figure}

In Fig.\,3c we show the histograms $P(E)$ for $p=0.06$.
While the two lattice orientations have 
the same energy distribution, there is a clear difference
between the liquid and lattice.  This results in a finite energy jump
$\Delta E$, and hence a finite entropy jump $\Delta E/T$, 
between the liquid and lattice.  The melting transition is 
clearly first order.  The histograms $P(Q)$ for $p=0.06$ are shown in 
Fig.\,3e, where we also see a finite jump $\Delta Q$ between liquid 
and lattice; for $Q$ the disorder also couples differently to the two lattice 
orientations.

In Fig.\,3d we show the histograms $P(E)$ for the more strongly 
disordered $p=0.09$, where only one lattice orientation is found.
In contrast to Fig.\,3c, we find that the energy distributions of 
the liquid and the lattice are now {\it identical!}  Thus 
there is no energy jump, and hence no entropy jump, and moreover 
no specific heat jump, in going from liquid to lattice.  However  
the histograms of $P(Q)$ plotted in Fig.\,3f remain clearly 
different for liquid and lattice, and so
there remains a finite jump $\Delta Q$ at the melting transition.  The 
transition at $p=0.09$ therefore remains first order, 
even though the entropy jump has vanished. Combining $\Delta E=0$
with the Clausius-Clapeyron relation, we conclude that the melting
line must now be perfectly parallel to the temperature axis \cite{R13}, 
and the transition becomes disorder, rather than thermally, driven.  As 
shown in Fig.\,2, we have been able to follow the 
melting line from where it first turns parallel to the $T$ 
axis, down to several lower temperatures. 

In Fig.\,4 we plot $\Delta E$ and $\Delta Q$ vs. $p$ along the melting 
line.  That $\Delta E$ and $\Delta Q$ never simultaneously vanish, 
indicates that no critical point exists along the melting line.

Although our simulations are for a specific fixed value of $B$, 
if we assume that the above result continues to hold 
for general values of $B$, then it must be true that the phase 
diagram in the $B-T$ plane at fixed $p$ must similarly turn parallel 
to the $T$ axis at low temperatures.  The point where the melting 
line first turns parallel to the $T$ axis has many features in common 
with an ``upper critical point''; discontinuities as $T$ varies 
across the melting line below this point, will cease to exist as $T$ 
varies above this point.  Yet there is no true critical end point and 
the first order melting line extends continuously down to lower 
temperatures.
\begin{figure}
\epsfxsize=3.2truein
\epsfbox{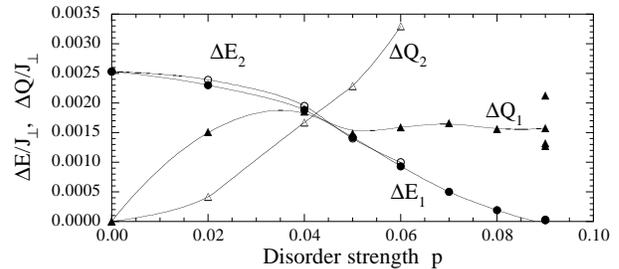}
\vspace{9pt}
\caption{
Jumps $\Delta E$ and $\Delta Q$, 
vs. $p$, along the melting line, $T_c(p)$, for $L_z=16$.
Subscripts ``1'' and ``2'' refer to the two possible orientations
of the vortex lattice (solid and open symbols respectively).
}
\label{f4}
\end{figure}

Next, we investigate the superconducting phase coherence in our 
model, by computing the helicity modulus \cite{R11}, $\Upsilon_z$, parallel 
to the applied magnetic field.
$\Upsilon_z=0$ indicates the absence of phase coherence.
In Fig.\,5a we show $\Upsilon_z$ vs. $T$, for two different disorder 
strengths $p<p_c=0.09$, comparing systems with  
$L_z=16$ and $L_z=24$.  In both cases we see a discontinuous jump in 
$\Upsilon_z$ at the melting transition (at $T_c$ we compute
$\Upsilon_z$ separately for the coexisting lattice and liquid phases, 
using the decomposition of our states according to $\Delta S_1$).  
As $L_z$ is increased, we see that $\Upsilon_z$ vanishes in the vortex liquid;
the vortex lattice melting thus marks the loss of superconducting phase 
coherence.  

\begin{figure}
\epsfxsize=3.2truein
\epsfbox{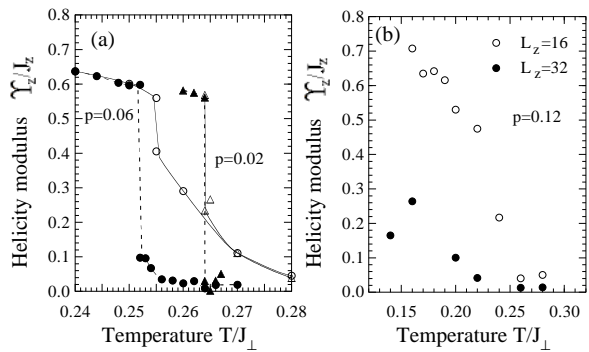}
\caption{
Longitudinal helicity modulus $\Upsilon_z/J_z$ vs. $T$ for disorder 
strengths (a) $p=0.02$ and $p=0.06<p_c$ (open symbols and solid lines are for 
$L_z=16$; solid symbols and dashed lines
are for $L_z=24$) and (b) $p=0.12>p_c$.
}
\label{f5}
\end{figure}
We now consider $p>p_c$ to see if a phase coherent vortex 
glass state might exist above the disorder driven melting of the 
vortex lattice.  In Fig.\,5b we show $\Upsilon_z$ vs. $T$ for 
$p=0.12$ and system sizes $L_z=16$ and $L_z=32$.  Although it becomes
extremely difficult to equilibrate our $L_z=32$ system at such
low $T$, we see a dramatic finite size effect strongly suggesting
that $\Upsilon_z$ decreases to zero as $L_z$ increases.  
We therefore find no evidence for a true phase coherent 
vortex glass.  
Our result agrees
with conclusions from dynamical simulations of an interacting vortex 
line model by Reichhardt {\it et al} \cite{R8}.
In contrast, recent work
by Kawamura \cite{R14} found a finite vortex glass $T_c$ in numerical studies of
a model similar to Eq.(\ref{eH}), but at a higher field density 
$f=1/2\pi$ and $J_{ij}$ uniformly 
distributed on $[0,2]$.  It is not clear if this 
disagreement is due his much stronger random pin strengths, or possible
finite size effects in his systems of $L\le 16$.  
A vortex glass transition has also recently been found for
an interacting vortex line model with $f=1/2$ and very strong pinning 
\cite{R15}.

Finally, in Fig.\,6 we plot the longitudinal phase 
angle correlation length, $\xi_z$,  in the liquid $T\ge T_c(p)$,
as determined by fitting the correlation function,
\begin{equation}
	C(z)=\sum_j\langle e^{i[\theta_j-\theta_{j+z\hat z}]}\rangle\enspace,
    \label{eC}
\end{equation}
to an exponential decay for $z<L_z/2$.  We see that for fixed $T$,
$\xi_z$ decreases as $p$ increases.  However, since $T_c(p)$ decreases
as $p$ increases, the value $\xi_z(T_c(p))$ at melting
{\it increases} as $p$ increases.  
Even in the disordered state above $p_c$ we find
that $\xi_z$ can be as large as in the liquid just above 
the thermally driven melting line.
This suggests that the disorder 
driven melting should not be thought of as a layer decoupling transition.
\begin{figure}
\epsfxsize=3.2truein
\epsfbox{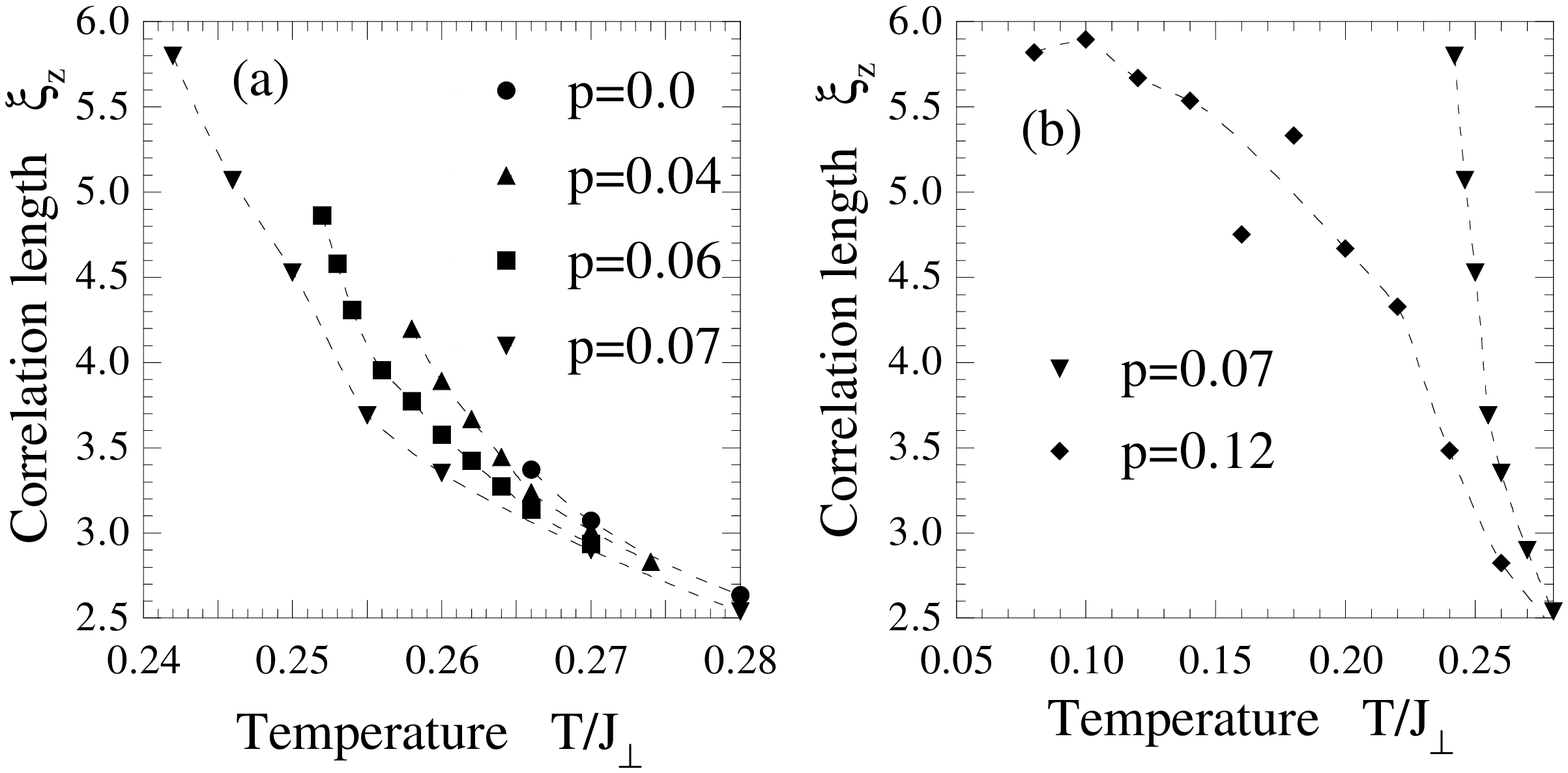}
\caption{
Longitudinal correlation length $\xi_z$ vs. $T$ in the vortex
liquid phase for (a) several values of disorder $p<p_c=0.09$
at $L_z=24$; and for (b) $p=0.12>p_c$ at $L_z=32$, in
comparison to $p=0.07<p_c$ from (a).
}
\label{f6}
\end{figure}

As we were finishing this work, we learned of similar work
by Nonomura and Hu \cite{R16}, using the same model (\ref{eH}) with a 
slightly different scheme for the point randomness and a much
weaker inter-planar coupling, $J_z/J_\perp =1/400$.  Using different
methods they too find a disorder driven 1st order vortex lattice melting 
line nearly parallel to the $T$ axis at low $T$.  However they also
claim to find a 1st order ``vortex slush'' to liquid transition 
extending to higher disorder from the thermally driven melting line, 
as well as a vortex glass to vortex slush transition at lower $T$.
We too find a peak in specific heat for $p>p_c$ that lies at a $T$
in the vicinity of the thermally driven melting line, however
we have interpreted this as a smooth crossover rather than a
true phase transition.

We would like to thank C. Marcenat for a valuable discussion.
This work was supported by the
Engineering Research Program of the Office of Basic Energy Sciences
at the Department of Energy grant DE-FG02-89ER14017, the
Swedish Natural Science Research Council Contract No. E 5106-1643/1999,
and by the resources of the Swedish High Performance Computing Center 
North (HPC2N).  Travel between Rochester and Ume{\aa} was supported by 
grants NSF INT-9901379 and STINT 99/976(00).


\begin{thebibliography}{99}
\bibitem{R13}
Considering $dF(T,p_c(T))/dT = \partial F/\partial 
T +NQdp_c/dT$ for the free energies of liquid and lattice phases,
and noting that along the melting line $F$ and $E\propto\partial 
F/\partial T$ are equal in the two phases while the $Q$ are 
not, implies that $dp_c(T)/dT=0$.
\end{thebibliography}
\end{document}